\documentclass[twocolumn,showpacs,preprintnumbers,amsmath,amssymb,superscriptaddress]{revtex4-1}
\pdfoutput=1

\usepackage{amssymb}
\usepackage{amsfonts}
\usepackage{amsmath}
\usepackage{graphicx}
\usepackage{textcomp}
\usepackage{mathptmx}
\usepackage{dcolumn}
\usepackage{bm}
\usepackage{times}
\usepackage{color}

\newcommand{\bra}[1]{\langle #1|}
\newcommand{\ket}[1]{|#1\rangle}
\begin{document}

\title{Experimental demonstration of a resonator-induced phase gate in a multi-qubit circuit QED system}

\author{Hanhee Paik}
\thanks{These authors contributed equally to the work.}
\author{A. Mezzacapo}
\thanks{These authors contributed equally to the work.}
\author{Martin Sandberg}
\thanks{These authors contributed equally to the work.}
\author{D. T. McClure}
\author{B. Abdo}
\author{A. D. C\'{o}rcoles}
\author{O.~Dial}
\affiliation{IBM T. J. Watson Research Center, Yorktown Heights, NY 10598-0218, USA}
\author{D. F. Bogorin}
\author{B. L. T. Plourde}
\affiliation{Department of Physics, Syracuse University, Syracuse, NY 13244-1130, USA}
\author{M. Steffen}
\author{A. W. Cross}
\author{J. M. Gambetta}
\author{Jerry M. Chow}
\affiliation{IBM T. J. Watson Research Center, Yorktown Heights, NY 10598-0218, USA}

\date{\today}

\begin{abstract}
The resonator-induced phase (RIP) gate is a multi-qubit entangling gate that allows a high degree of flexibility in qubit frequencies, making it attractive for quantum operations in large-scale architectures. We experimentally realize the RIP gate with four superconducting qubits in a three-dimensional (3D) circuit-quantum electrodynamics architecture, demonstrating high-fidelity controlled-Z (CZ) gates between all possible pairs of qubits from two different 4-qubit devices in pair subspaces. These qubits are arranged within a wide range of frequency detunings, up to as large as 1.8 GHz. We further show a dynamical multi-qubit refocusing scheme in order to isolate out 2-qubit interactions, and combine them to generate a four-qubit Greenberger-Horne-Zeilinger state.
\end{abstract}


\maketitle

As recent progress in superconducting quantum processors has marched towards more complex networks of qubits~\cite{Kelly15a, Riste15a, Corcoles15a}, it becomes increasingly crucial to develop robust protocols for multi-qubit control. In particular, there has been a considerable amount of work aimed at improving single-~\cite{Sheldon16a} and two-qubit~\cite{Corcoles13a, Barends14a, Sheldon16b} controls for superconducting transmon devices. Although the fidelity of single-qubit gates ($>0.999$) has already been pushed above fault-tolerant thresholds for error correction codes such as the surface code~\cite{Bravyi98a, Fowler12a}, the study of two-qubit gates in multi-qubit systems is still an area of great exploration.

Currently, many two-qubit gates for superconducting qubits require specific arrangements of qubit frequencies to perform optimally. For example, the dynamically-tuned controlled-Z (CZ) gate~\cite{Strauch03a, DiCarlo09a, Barends14a} functions through magnetic flux-tuning two qubits into a specific resonance condition involving higher energy levels, which will not work if any other existing energy levels intervene between the qubits. A similar limitation arises with the all-microwave cross-resonance (CR) gate~\cite{Paraoanu06a, Rigetti10a, Chow11a, Sheldon16b}. The CR gate works for qubits within a narrow window of detunings defined by the anharmonicity of the qubit~\cite{Magesan16a}. This restriction becomes accentuated in larger networks of qubits where all qubit frequencies must be arranged within a small frequency window~\cite{Takita16a}.

\begin{figure}
\includegraphics[width=3.3in]{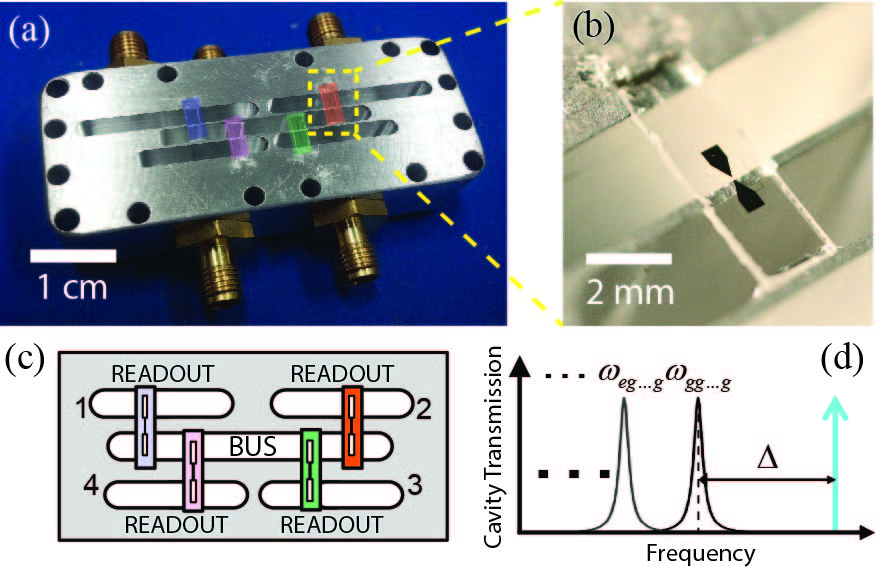}
\caption{\label{fig1} (a) Picture of our superconducting 4-qubit 3D cQED system with 5 cavities. The cavity enclosure is machined out of 6061 aluminum and connectorized by non-magnetic SMA feedthroughs. Four qubit chips (false colored: red, blue, green and pink) are mounted to couple to the bus cavity (center pocket) and individual readout cavities (outer pockets). (b) Close-up photograph of a 3D qubit chip mounted in the 5-cavity enclosure. (c) Diagram of the 4-qubit 3D cQED system with 5 cavities. Each qubit has an individual readout cavity (4 outer pockets). (d) Illustration of bus cavity transmission. The microwave drive for the RIP gate (cyan arrow) is blue-detuned by frequency $\Delta$ from the dressed cavity resonance $\omega_{gg...g}$. }
\end{figure}

\begin{figure*}
\includegraphics[width=7in]{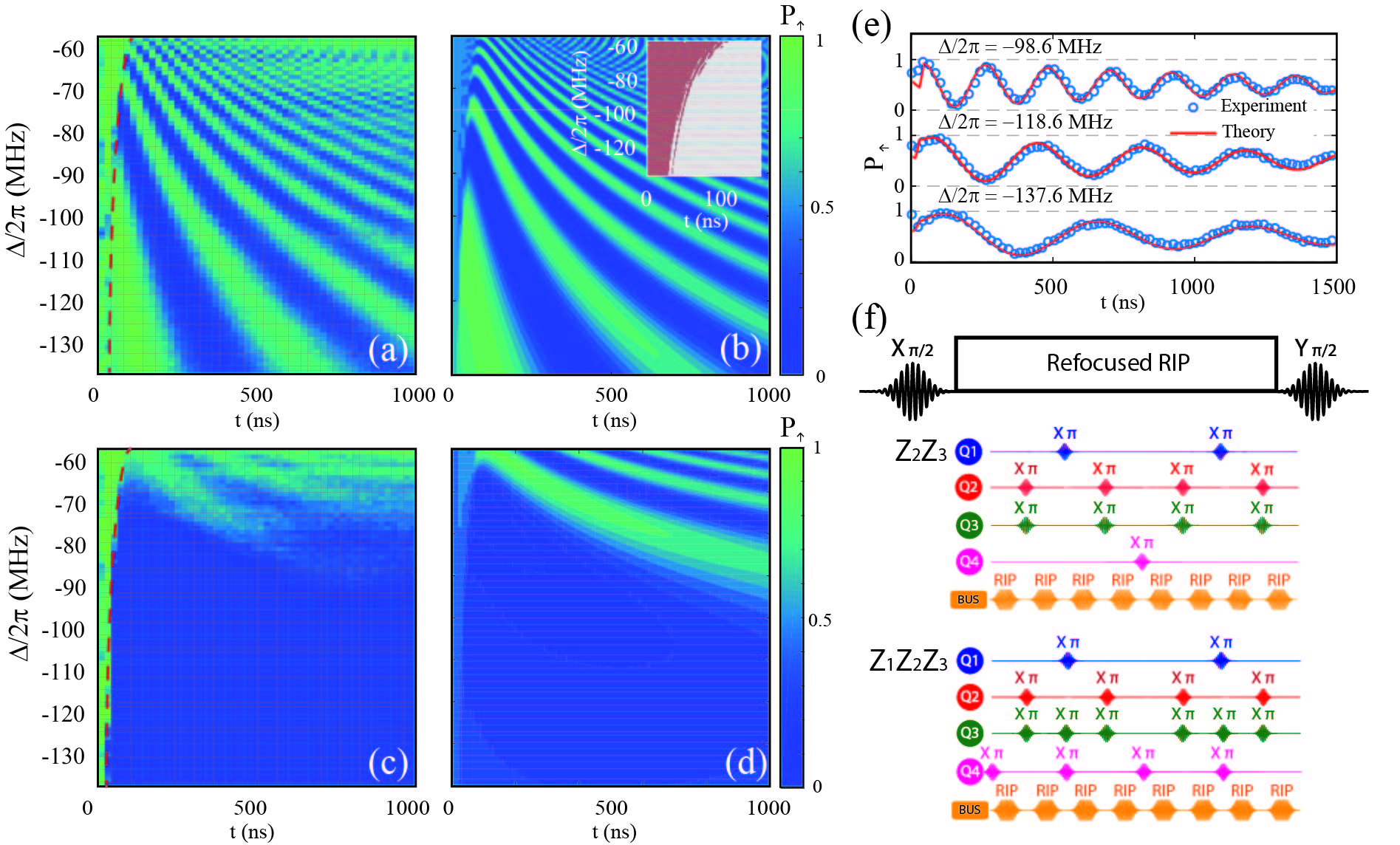}
\caption{\label{fig2} (a) Excited state population of the qubit $A$2 (see Table I) versus single RIP pulse gate time $t$ and detuning $\Delta/2\pi$, measured using a Ramsey experiment with a refocused RIP gate scheme for $Z_2Z_3$ in (f). The plot shows ZZ interaction between qubits $A$2 and $A$3. The red dashed line indicates a threshold gate time $\propto 1/\Delta$, below which no coherent oscillation is observed. (b) Theoretical prediction of the driven ZZ oscillations at the drive amplitude $\tilde{\epsilon}_R/2\pi=315$~MHz. [Inset] Residual photons versus single RIP pulse gate time $t$ and detuning $\Delta/2\pi$. The red region is where the residual photons $>$ 0.01; the number of residual photons drops sharply after the threshold time. In our pulse shape, the rise time decreases as the gate time decreases, which causes the non-adiabatic drive. (c) Excited state population of $A$2 from the Ramsey experiment with with a refocused RIP gate scheme pulse sequence for $Z_1Z_2Z_3$ in (f), showing ZZZ interactions among qubits $A1$, $A2$ and $A3$ as a function of single RIP pulse gate time $t$. (d) Theoretical calculation of ZZZ interactions at amplitude $\tilde\epsilon_R/2\pi=200$~MHz. (e) Excited state population ($P_\uparrow$) of $A$2 versus single RIP pulse gate time $t$ at three detuning points. ZZ oscillations measured from the Ramsey experiments (blue circles) and calculated from the theory (red curves) show good agreement. The theoretical drive amplitude is fine-tuned at $\tilde{\epsilon}_R/2\pi=315\pm15$~MHz for the three rounds of measurements. (f) Pulse sequences for the Ramsey experiment (top) and 4-qubit refocused RIP gate schemes. To obtain ZZ and ZZZ interactions shown in (a)-(e), the Ramsey experiment is performed as a function of the drive detuning $\Delta$ while applying an 4-qubit refocused RIP gate scheme~\cite{Suppl} that singles out the $Z_2Z_3$ (middle) or $Z_1Z_2Z_3$ (bottom).}
\end{figure*}

A notable advantage of the resonator-induced phase (RIP) gate~\cite{Cross15a, Puri2016} is its capability to couple qubits even if they are far detuned from each other. Therefore, the RIP gate can overcome difficulties due to constraints on the frequency arrangements of the qubits that can hinder scalability towards larger quantum architectures. The RIP gate is a CZ gate that exploits strong coupling between qubits and a resonator in a circuit quantum electrodynamics (cQED) system. It is realized by applying a detuned pulsed microwave drive to a shared bus cavity, without a strong requirement on the qubit frequencies. In addition, it is insensitive to phase fluctuations of the drive, depending only on drive amplitude and detuning.

In this Letter, we experimentally demonstrate the RIP gate in two cQED devices, each of which composed of four three-dimensional (3D) transmon superconducting qubits~\cite{Paik11a} coupled to both a central bus cavity and individual readout cavities. First, we show that a variety of state-dependent phases are induced by the RIP gate. Phases that originate via weight two and three Pauli operators are singled out using echo sequences and measured. Our experiments confirm the predicted dependence of the acquired phases on drive amplitude and detuning. Then, we demonstrate the frequency flexibility of the RIP gate by performing the gate between 12 individual qubit pairs from two devices, with qubit-qubit detunings up to 1.8 GHz. High-fidelity CZ gates are observed in pair subspaces using two-qubit randomized benchmarking. Finally, using pairwise CZ interactions in the four-qubit subspace, we generate a 4-qubit Greenberger-Horne-Zeilinger (GHZ) state.

Figure~\ref{fig1}(a) shows our 4-qubit 3D cQED device. The qubits are placed on individual 2 mm$\times$6 mm HEM sapphire \cite{Dial15a} chips [see Fig.1(b)], mounted in a 5-cavity enclosure machined from 6061 aluminum. The parameters of the two 4-qubit devices, labeled as Device $A$ and Device $B$, are listed in Table~\ref{table1}. Experimental setup details are given in the Supplemental Material~\cite{Suppl}. The qubit states are measured via low-power dispersive readout. The Josephson parametric converters (JPC)~\cite{Abdo11a} are only used on qubits $B$1 and $B$2 (see Table I) and the rest of qubits are measured without JPCs. The single-qubit and simultaneous randomized benchmarking results show that all single-qubit gate fidelities are higher than 0.999~\cite{Suppl}, confirming that we do not have any significant addressability errors~\cite{Gambetta12a}.

\setlength{\belowcaptionskip}{3pt}

\begin{table*}[t]
\caption{\label{table1} Parameters of the two 4-qubit devices used in the experiments (Device $A$ and Device $B$). $\omega_q$ is the qubit frequency, $\delta$ is the qubit anharmonicity, $\chi$ is the qubit-bus dispersive frequency shift, $\omega_c$ is the readout cavity frequency, $T_1$ is the energy decay time of the qubit, $T_{2}^*$ is the Ramsey coherence time and $T_{echo}$ is the Hahn echo coherence time. Values of $T_1$, $T_{echo}$, and $T_{2}^*$ that were measured multiple times during the experiment, are listed as a range in the Table. The bus cavity frequencies are 6.9676 GHz (Device $A$) and 6.9710 GHz (Device $B$). In both cases, the bus cavity has a decay rate $\kappa/2\pi$ = 7.7 kHz. Qubits are labeled according to their locations in the cavity enclosure as shown in Fig. 1(c).}
\begin{ruledtabular}
\begin{tabular}{cccccccc|cccccccc}
Qubit & $\omega_q/2\pi$   & $\delta/2\pi$ & $\chi/2\pi$ & $\omega_c/2\pi$ & $T_1$ & $T_{echo}$ & $T_{2}^*$ &
Qubit & $\omega_q/2\pi$   & $\delta/2\pi$ & $\chi/2\pi$ & $\omega_c/2\pi$ & $T_1$ & $T_{echo}$ & $T_{2}^*$ \\
Index & (GHz) & (MHz) & (MHz) & (GHz) & ($\mu s$) & ($\mu s$)& ($\mu s$) &
Index & (GHz) & (MHz) & (MHz) & (GHz) & ($\mu s$) & ($\mu s$)& ($\mu s$) \\
\hline
$A1$ & 5.7862   & 305 & 10    & 10.2020 & 26-36 & 36-40& 6-17  & $B1$ & 5.7828 & 303 & 6.8 & 10.1949 & 31-36 & 30-40  & 27  \\
$A2$ & 5.1459   & 304 & 3.7   & 10.0846 & 63-68 & 49-68& 21-23 & $B2$ & 4.5597 & 287 & 0.7  & 10.0805 & 88-90 & 46-86 & 48  \\
$A3$ & 6.3037   & 243 & 4.6   & 9.9799  & 45-59 & 22-34& 12-42 & $B3$ & 6.3657 & 234 & 6.7  & 9.9775  & 46-59 & 16-27 & 12  \\
$A4$ & 4.7630   & 280 & 2.2   & 9.8328  & 56-68 & 45-46& 37    & $B4$ & 4.9624 & 284 & 0.1  & 9.8553  & 38-45 & 33-36 & 18  \\
\end{tabular}
\end{ruledtabular}
\end{table*}

Our four-qubit system is described by a sum of Duffing oscillator Hamiltonians coupled to the bus cavity, with a microwave drive term for the RIP gate~\cite{Suppl}. If the qubit frequencies are sufficiently spaced, the qubit-qubit interactions become diagonal in the qubit computational basis, with a static component and dynamical interactions activated by the cavity drive. The qubit interactions can therefore be described in terms of $Z$ operators, which makes the RIP gate insensitive to phase fluctuations in the drive. To grasp the different interactions involved in the dynamics, we compute the phase accumulation rate $\dot{\theta}$ from the qubit interactions at the steady state under the action of an unmodulated drive $\tilde{\epsilon}(t)=\epsilon_I(t)+i \epsilon_Q(t)=\tilde{\epsilon}_0$,
\begin{eqnarray}
\dot{\theta}_{Z_iZ_j}&=&-\frac{|\tilde{\epsilon}_0|^2\chi^2}{8 \Delta(\Delta+2\chi)(\Delta+4\chi)},\label{theta1}\\
\dot{\theta}_{Z_aZ_bZ_c}&=&-\frac{3|\tilde{\epsilon}_0|^2\chi^3}{16\Delta(\Delta+\chi)(\Delta+3\chi)(\Delta+4\chi)},\label{theta2}\\
\label{theta3b}
\dot{\theta}_{Z_1Z_2Z_3Z_4}&=&-\frac{3|\tilde{\epsilon}_0|^2\chi^4}{8\Delta(\Delta+\chi)(\Delta+2\chi)(\Delta+3\chi)(\Delta+4\chi)},
\end{eqnarray}
where $\Delta$ is the detuning of the drive frequency to the dressed bus cavity with all qubits in the ground state. We have assumed that each qubit has the same dispersive shift $\chi$. Eqs.~(\ref{theta1}-\ref{theta3b}) reveal a scaling of the diagonal interactions in drive amplitude $|\tilde{\epsilon}_0|$ and detuning 1/$\Delta$ as well as $(\chi/\Delta)^p$ with the increasing Pauli weight $p$. A single RIP gate tone will turn on all Z interactions at the same time. However, with the nominal condition $\chi/\Delta < 1$, the multi-body interaction rate becomes slower as the Pauli weight increases.

To observe the amplitude- and frequency-scaling behavior of the phases from weight two and three Z operators, we perform a series of Ramsey experiments while applying the RIP gate. Refocused RIP gate schemes are designed for 4 qubits to single out pairwise ZZ or ZZZ terms, and the pulse sequences are shown in Fig.~\ref{fig2}(f). In the refocused RIP gate scheme, $X_{\pi}$ pulses on each qubit are applied between RIP gate pulses, and echo away unwanted Z interactions of various Pauli weights. For the RIP gate pulses, we use an adiabatic drive of the form $\tilde{\epsilon}_R(t)=\tilde{\epsilon}_A(1+\cos(\pi \cos(\pi t/\tilde{\tau}))$ \cite{Paik13a}, where $\tilde{\tau}$ is the pulse width. This pulse shape suppresses the photon population of the cavity to third order in the cavity-drive detuning~\cite{Cross15a}.

Examples of two-qubit interactions ($Z_2Z_3$) between qubit $A2$ and $A3$, and three-qubit interactions ($Z_1Z_2Z_3$) between qubits $A1$, $A2$ and $A3$ from Device $A$ are shown in Fig.~\ref{fig2}(a) and Fig.~\ref{fig2}(c) respectively, as a function of a single RIP gate pulse width $t$ and the detuning $\Delta$. Both $Z_2Z_3$ and $Z_1Z_2Z_3$ become faster as $\Delta$ approaches zero, as predicted by the steady-state solution of Eqs.~(\ref{theta1}-\ref{theta2}). The experimental $Z_2Z_3$ and $Z_1Z_2Z_3$ are compared with a closed form solution for the density matrix of the system~\cite{Suppl}. We find excellent agreement between the experiment [Fig.~\ref{fig2}(a) and Fig.~\ref{fig2}(c)] and theory [Fig.~\ref{fig2}(b) and Fig.~\ref{fig2}(d)], observing deviations of $\sim 0.5\%$ in the pulse amplitude for the different drive detunings. Fig.~\ref{fig2}(e) shows $Z_2Z_3$ from three different drive detunings. The deviations are likely related to cavity nonlinearity and frequency-dependent attenuation of the drive lines.

\setlength{\belowcaptionskip}{3pt}
\begin{table}
\caption{\label{tab:fid} RIP gate fidelities measured on all 12 different qubit pairs in Device $A$ and $B$. The detuning between the bus cavity and the RIP gate drive is 40 MHz in all 12 experiments. $\Delta_q$ is the detuning between control and target qubits, $T_{\rm gate}$ is the total gate time including two RIP gate pulses and a single qubit echo pulse (a 36.7 ns wide $\pi$ pulse) shown in Fig.~\ref{fig3}, $\zeta$ is the rate of static ZZ interaction, $F_{\rm coh}$ is the coherence limit on the gate fidelity estimated based on the total gate time and worst $T_1$ and $T_{\rm echo}$, $F_C$ stands for the fidelity per Clifford from the two-qubit randomized benchmarking (RB) and $F_g$ is the fidelity per CZ generator estimated from the average number ($N_C$) of generators per Clifford using $F_g = 1- (d-1)(1-\alpha^{1/N_C})/d$ with $d = 2^n$, $N_C$ = 1.5. $\alpha$ is the exponent of the decay model from two-qubit randomized benchmarking.}
\begin{ruledtabular}
\begin{tabular}{ccccccc}
$\Delta_q/2\pi$ & Qubit &$T_{\rm gate}$ & $\zeta/2\pi$    & $F_{\rm coh}$ & $F_C$  &   $F_g$     \\
(GHz)           & Pair  &   (ns)        &  (kHz)          &               &        &              \\
\hline
0.383      & $A2$-$A4$ & 525       &   60      &   0.9893   & 0.9577(6)  & 0.9787(3)  \\
0.403      & $B2$-$B4$ & 760       &   10      &   0.9832   & 0.9320(12) & 0.9655(6)   \\
0.518      & $A1$-$A3$ & 285       &   138     &   0.9913   & 0.9665(9)  & 0.9831(4)  \\
0.583      & $B1$-$B3$ & 472       &   156     &   0.9772   & 0.9554(6)  & 0.9775(3)  \\
0.637      & $A1$-$A2$ & 285       &   107     &   0.9883   & 0.9683(9)  & 0.9841(4)  \\
0.820      & $B4$-$B1$ & 472       &   34      &   0.9827   & 0.9501(7)  & 0.9748(4)   \\
1.023      & $A4$-$A1$ & 461       &   60      &   0.9857   & 0.9532(8)  & 0.9764(4)  \\
1.158      & $A2$-$A3$ & 413       &   60      &   0.9861   & 0.9709(7)  & 0.9853(3) \\
1.223      & $B1$-$B2$ & 424       &   16      &   0.9872   & 0.9651(7)  & 0.9825(3)  \\
1.403      & $B3$-$B4$ & 424       &   10      &   0.9805   & 0.9486(4)  & 0.9741(2)  \\
1.541      & $A3$-$A4$ & 509       &   30      &   0.9824   & 0.9674(6)  & 0.9836(3)  \\
1.806      & $B2$-$B3$ & 424       &   23      &   0.9831   & 0.9670(6)  & 0.9834(3)  \\
\end{tabular}
\end{ruledtabular}
\end{table}

In our experiment, the rise time of the RIP gate pulse shortens as the pulse width decreases in the pulse shape $\tilde{\epsilon}_R(t)$. As a result, both the experiment and theory in Fig.~\ref{fig2} reveal a threshold time, below which the gate is strongly inhibited due to non-adiabatic driving. A fast rise of the RIP gate pulse is signalled by the presence of residual photons in the bus at the end of the gate. The non-adiabatic time-threshold is marked with a red dashed line in Fig.~\ref{fig2}(a) and (c), and is inversely proportional to the drive detuning $\Delta$. The closed-form solutions~\cite{Suppl} indicate a finite amount of residual photons $\langle n(t)\rangle >0.01$ for short gate times, as plotted as the red region in the inset of Fig.~\ref{fig2}(b).
\setlength{\belowcaptionskip}{-10pt}
\begin{figure}
\includegraphics[width=3.6in]{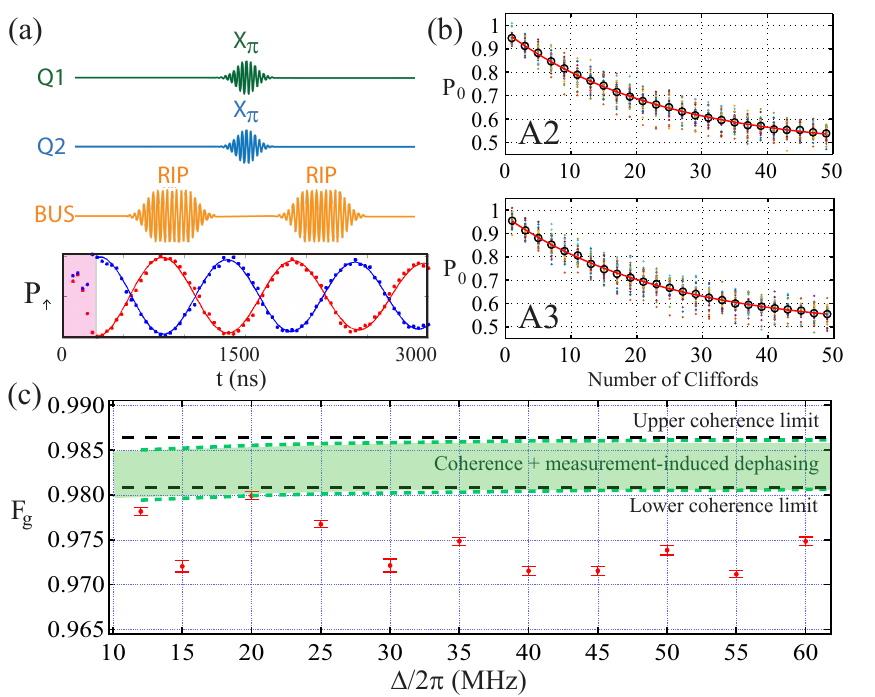}
\caption{\label{fig3} (a) Pulse sequence of the two-qubit refocused RIP gate scheme and a plot of the excited state population (P$_\uparrow$) of a target qubit measured from a RIP tune-up procedure using a Ramsey experiment. The blue and red curves show Ramsey oscillations from a target qubit when a control qubit is in the ground state (blue) and in the excited state (red). The non-adiabatic region is highlighted with pink. (b) Population of $|0\rangle$ vs the number Cliffords measured from the two-qubit Clifford randomized benchmarking on $A2$-$A3$ at the drive detuning $\Delta/2\pi$ = 20 MHz. Forty different randomization sequences are generated and applied in the experiment (colored dots). (c) $ZZ_{\pi/2}$ fidelity between $A2$-$A3$ versus the RIP gate drive detuning $\Delta/2\pi$ (red dots), showing no appreciable dependence on $\Delta$. The error bars are obtained from the fit from two-qubit RB. Theoretically calculated upper- and lower-limits on the gate fidelity without (black dashed line) and with the measurement-induced dephasing present in the system (green region/dashed curves) are plotted together. The upper- and lower-limits are calculated from measured minimum and maximum coherence times in Table~\ref{table1}.}
\end{figure}

To demonstrate the flexibility of the RIP gate with respect to qubit frequencies, we characterize the gate performance via two-qubit randomized benchmarking (RB) ~\cite{Magesan11a, Corcoles13a} over a large range of qubit-qubit detuning $\Delta_q$. For the characterization, we restrict our experiment to a two-qubit subspace with other two qubits in the ground state. The two-qubit refocused RIP gate scheme~\cite{Chow13a}, illustrated in Fig.~\ref{fig3}(a), is used to realize a two-qubit CZ generator $ZZ_{\pi/2} =$ exp$[-i\frac{\pi}{4}\sigma^{Z}_{l}\otimes\sigma^{Z}_{m}]$. The gate is tuned up using Ramsey experiments shown in Fig.~\ref{fig2}(f), by performing first a $X_{\pi/2}$ gate on the target qubit, then applying the refocused RIP gate with a varying gate time. The final $Y_{\pi/2}$ on the target qubit ensures, when the phase of the target qubit is $\pm\pi/2$, a maximal contrast between two Ramsey curves for each control qubit state. The lower plot in Fig.~\ref{fig3}(a) shows two out-of-phase Ramsey curves as a function of gate time from a tune-up procedure. In the tune-up, the minimum gate time is typically bounded by the non-adiabatic time-threshold, which is highlighted in pink in Fig.~\ref{fig3}(a).

High fidelity $ZZ_{\pi/2}$ interactions are achieved by the RIP gate between all qubit pairs, up to 1.8 GHz in qubit-qubit detuning. The fidelity data is summarized in Table~\ref{tab:fid}. The fidelity per Clifford ranges from 0.93 to 0.97, corresponding to 0.96 to 0.98 fidelity per CZ generator (see Table~\ref{tab:fid} caption). We find that the effect of measurement-induced dephasing \cite{Gambetta06a} is small in our devices. The measurement-induced dephasing is investigated by measuring the gate fidelity while varying the detuning of the RIP gate drive. The measurement-induced dephasing is expected to worsen as the detuning decreases. The RIP gate pulse width is fixed at 266.7 ns (total refocused RIP gate time = 570 ns) and we keep the gate time constant by adjusting the drive amplitude for all drive detuning. Fig.~\ref{fig3}(b) shows a two-qubit randomized benchmarking result from the qubit pair $A2$ and $A3$ at the drive detuning $\Delta/2\pi$ = 20 MHz. At the 20 MHz detuning, the fidelity per Clifford is 0.9709(7), corresponding to 0.9853(3) for the fidelity per CZ generator, which is close to the lower fidelity bound imposed by the coherence times and the measurement-induced dephasing. We find no appreciable dependence on the detuning of the RIP drive down to 12 MHz as shown in Fig.~\ref{fig3}(c), and below 12 MHz, the gate does not work due to non-adiabaticity. The overall RIP gate fidelity ($\sim$ 0.97) is close to the coherence limit (0.980 $\sim$ 0.985) and the estimated error from the measurement-induced dephasing is about 10$^{-3}$ at the lowest detuning of 12 MHz.

\begin{figure}
\vspace{0.1in}
\includegraphics[width=3.6in]{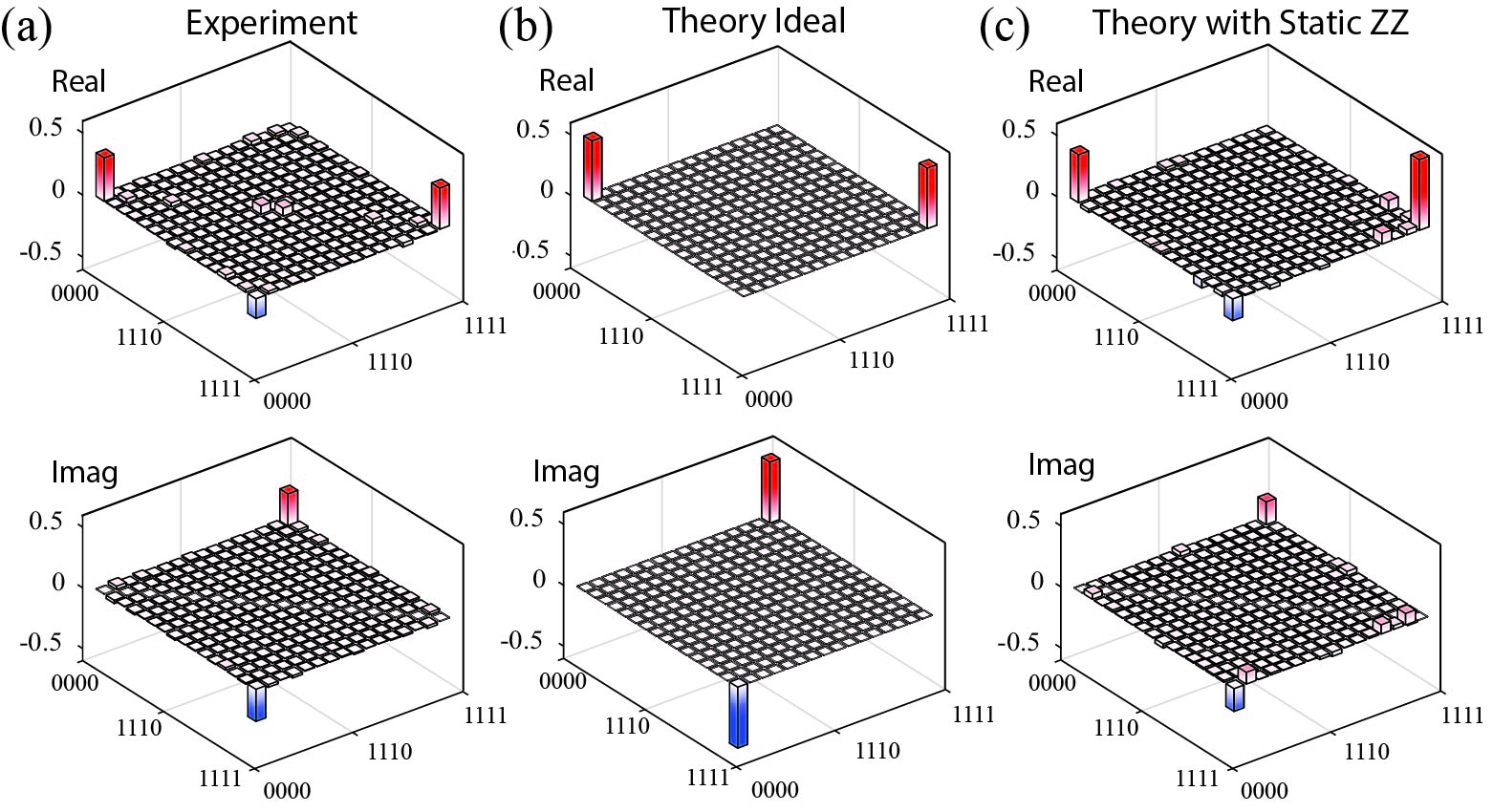}
\caption{\label{fig4}(a) Experimentally reconstructed density matrix of a 4-qubit GHZ state from the quantum state tomography performed on Device $A$. Quantum state tomography is performed using the method introduced in Ref.~\cite{Ryan15a}. The 4-qubit refocused RIP gate scheme is used to create the GHZ state. The quantum state fidelity is 60.5 $\%$ to the ideal GHZ state. (b) Theoretically reconstructed ideal density matrix of a 4-qubit GHZ state using the exact gate sequences from the experiment in (a). (c) Theoretical density matrix of a 4-qubit GHZ state. Static ZZ interactions are included at each single-qubit operation. }
\end{figure}

We implement 4-qubit refocused RIP gate schemes [Fig.~\ref{fig2}(f)] to perform pairwise CZ gates in the four-qubit space~\cite{Suppl}. We generate a maximally entangled 4-qubit GHZ state ($\ket{\Psi}=1/\sqrt{2}(\ket{0000}-i\ket{1111})$) in Device $A$ using CZ gates between the qubit pairs $A1$-$A2$, $A2$-$A3$ and $A3$-$A4$. Single RIP pulse widths are 203 ns for $A1$-$A2$ and $A2$-$A3$, and 173 ns for $A3$-$A4$, which makes the total CZ gate times to be 1.871 $\mu$s for $A1$-$A2$ and $A2$-$A3$, and 1.631 $\mu$s for $A3$-$A4$ with the single-qubit pulse width of 36.7 ns. The resulting GHZ density matrix is shown in Fig.~\ref{fig4}(a). The state fidelity to the ideal GHZ state [Fig.~\ref{fig4}(b)] is 60.5$\%$ with a maximum likelihood estimation, which is partly limited by decoherence during the long gate, and imperfect tuning of the 4-qubit refocused RIP gate scheme. We find that static ZZ interactions can model some of the non-ideality observed in the experiment, producing erroneous components in the density matrix. With this mode, the non-ideal matrix components in the experimental density matrix are reproduced in the theoretical density matrix shown in Fig.~\ref{fig4}(c).

In summary, we have implemented the all-microwave RIP gate in 4-qubit superconducting 3D cQED systems. The RIP gate induces ZZ interactions, which are insensitive to any phase fluctuations in the drive, easing requirements on phase stability for the qubit microwave controls. We have characterized 12 two-qubit CZ gates amongst a wide range of frequencies, spanning up to 1.8 GHz, and demonstrated high-fidelity. This flexibility in qubit frequencies and the demonstrated high fidelity make the RIP gate an attractive tool for quantum operations in a large-scale architecture.

We thank Easwar Magesan, Jim Rozen and Sarah Sheldon for experimental discussions and Markus Brink, George Keefe, and Mary Beth Rothwell for device fabrication. We acknowledge support from IARPA under contract W911NF-10-1-0324.

\vspace{3cm}
\begin{widetext}
\section{Supplemental Material for \\
Experimental demonstration of a resonator-induced phase gate in a multi-qubit circuit QED system}

\subsection{Experiment Setup}

The experimental setup is shown in Fig.~\ref{msmtsetup} as a block diagram.  Our input microwave signal is transmitted through the CuNi coax with a 10 dB attenuator at 50 K and 4 K, 6 dB at the still stage, 10 dB at the 100 mK stage and 20 dB at the mixing chamber stage (10 mK). The device is placed inside a Cryoperm cylinder. The output microwave signal, which is transmitted through two Quinstar CWJ isolators and a K$\&$L low-pass filter, goes through a NbTi coax cable from 10 mK to 4 K, and is amplified by a low-noise HEMT amplifier from Caltech or Low-Noise Factory at the 4 K stage. The output signal is further amplified by 8 - 12 GHz B$\&$Z low-noise amplifiers at room temperature.  For the readout, we used the low-power dispersive readout with heterodyne detection. For B1 and B2 in Device $B$, JPC quantum limited amplifiers are used. The readout signal is mixed down to a 12 MHz IF signal using a Marki image rejection mixer.

We use Holzworth 4-channel microwave sources (HS9004A) for single-qubit gates and cavity readout pulses, which are modulated by a BBN Arbitrary Pulse Sequencer with IQ mixers from PolyPhase. An Agilent E8267D is used for the RIP gate, which is modulated by a Tektronix 5014C. The single-qubit gates and readout pulses are single-sideband (SSB) modulated to avoid microwave leakage at the carrier frequency. The SSB frequencies are 50 MHz for single-qubit gates and 12 MHz for readout.  Carrier leakage and skewness were minimized by adjusting the dc offset and amplitude imbalance between the I and Q port of the mixer. The microwave signals for single-qubit gate operations are amplified by a Mini-Circuits amplifier, ZVA-183. All microwave generators are phase-locked to a 10 MHz rubidium frequency standard (SRS FS725).

\begin{figure*}
\includegraphics[width=6in]{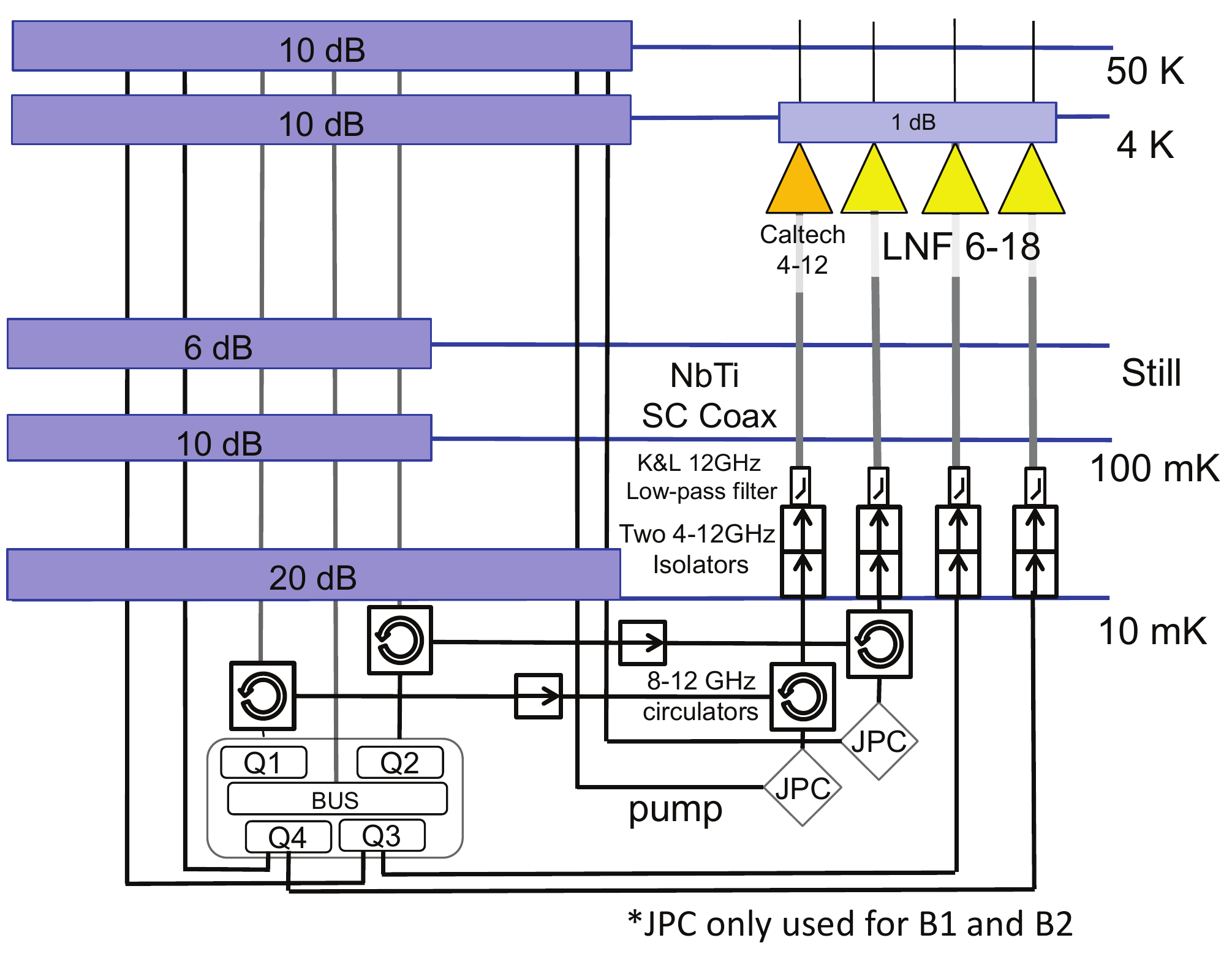}
\caption{\label{msmtsetup} Block diagram of the measurement setup. Experiments are performed in a Bluefors dilution fridge at 10 mK. Q1 and Q2 are measured in reflection, and Q3 and Q4 are measured in transmission. JPC amplifiers are only used for $B$1 and $B$2 in Device $B$.}
\end{figure*}

\subsection{Single- and Simultaneous Randomized Benchmarking}
\begin{figure*}
\includegraphics[width=6in]{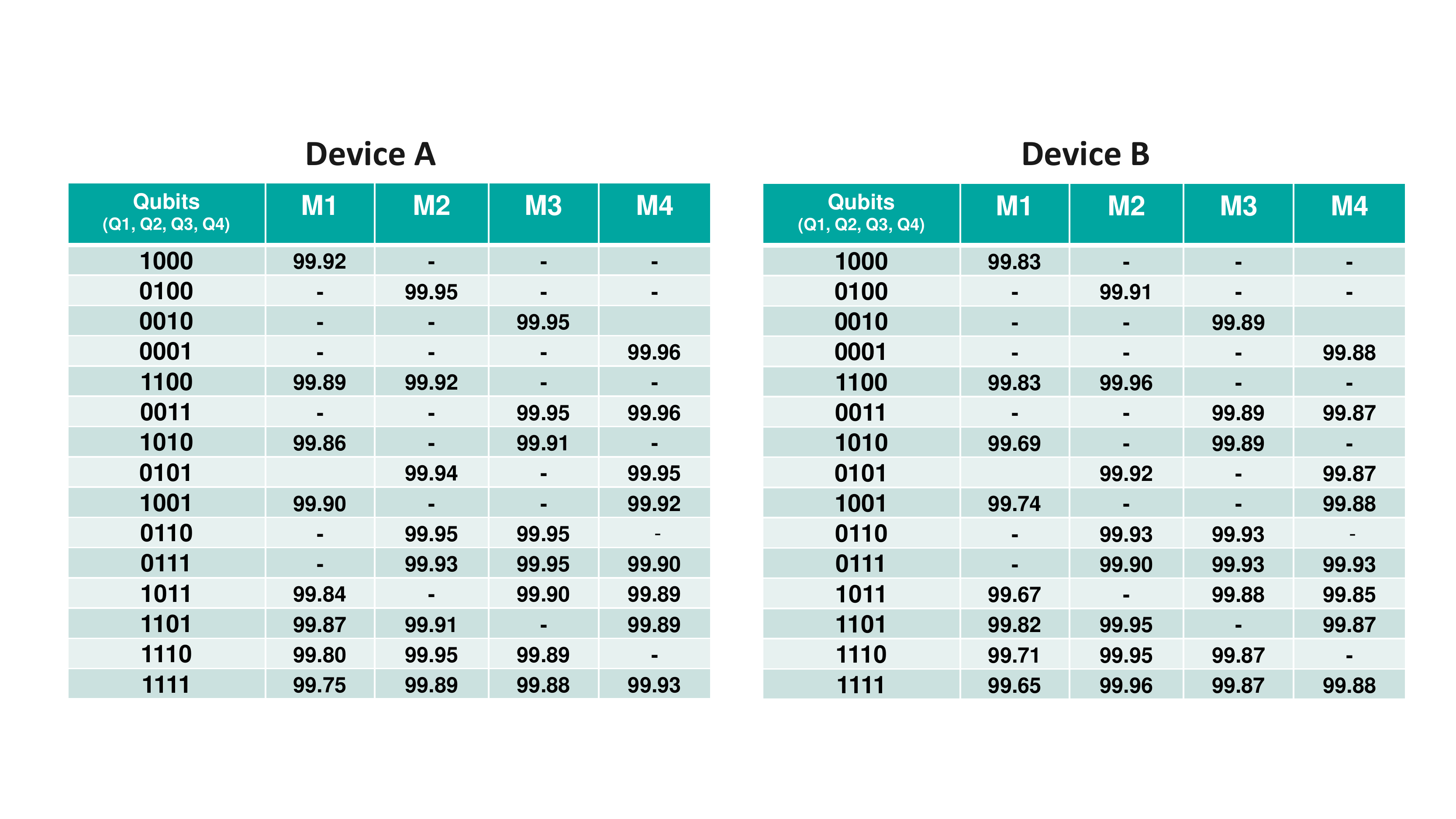}
\caption{\label{SSRB} Tables of single-qubit and simultaneous randomized benchmarking (RB) results. The ``1'' in the first column indicate qubits to which RB sequences are applied. Measurement results obtained from qubit $Q_i$ are denoted $M_i$.}
\end{figure*}
A summary of the single-qubit and simultaneous randomized benchmarking results from Device $A$ and $B$ is shown in Fig.~\ref{SSRB}.  The single-qubit gate is a 36.7 ns-long Gaussian pulse which is optimized with the DRAG technique~\cite{Motzoi09a}. The single-qubit gate fidelities are around 99.9$\%$ for for Device $A$ and 99.7 $\% \sim$ 99.9 $\%$ for Device $B$. In both qubits, we observe no visible degradation in fidelity in simultaneous benchmarking results, compared to single-qubit benchmarking, which indicates that there is no significant addressability error.

\subsection{Model for multi-qubit transmon systems}

We model our system composed of four 3D transmon qubits coupled to a common bus cavity by extending the two-qubit model presented in Ref.~\cite{Cross15aS} to a multi-qubit setup. We introduce the Duffing oscillator Hamiltonian that models the $i$-th qubit (here and in the following we set $\hbar=1$),
\begin{equation}
h_i=\omega_i a^\dag_i a_i +\frac{\delta_i}{2}a^\dag_i a_i(a^\dag_ia_i-1),
\end{equation}
where we have defined the bare transmon frequencies and anharmonicities~\cite{Koch07S} $\omega_i,\delta_i$, for the $i$-th qubit, together with the  excitation raising and lowering operators $a^\dag_i,a_i$. The total Hamiltonian for the interacting system of many qubits coupled to a single bus cavity can be then modeled as
\begin{equation}
H=\omega_r c^\dag c + \sum_{i} h_i + \sum_{i} g_i (a^\dag_i c+ c^\dag a_i),
\end{equation}
where $g_i$ is the coupling of the $i$-th qubit to the common cavity, $c^\dag$, $c$ are the cavity raising and lowering operators. If all the coupling terms $g_i$ and the anharmonicities $\delta_i$ are small compared to the transmon-cavity transition frequencies, $|\omega_i-\omega_r|\gg g_i,\delta_i$, then one can define an approximate dispersive Hamiltonian for the system
\begin{equation}
\label{DressedHam}
H_d=\omega_r c^\dag c +\sum_{i} h'_i +\sum_{i<j}\sum_{l,m}\sqrt{(l+1)(m+1)} J_{ij}^{(lm)}\left(\ket{l_i,m_j+1}\bra{l_i+1,m_j}+\ket{l_i+1,m_j}\bra{l_i,m_j+1}\right),
 \end{equation}
where we have introduced the notation $\ket{l_i,m_j}$, which stands for the $i$-th level for $l$-th transmon and the $j$-th level for the $m$-th transmon. The qubit Hamiltonians $h'_i$ include a dressing of the qubit frequency and a Stark shift term
\begin{equation}
h'_i=\sum_{k_i} \tilde{\omega}_{k_i} \ket{k_i}\bra{k_i}+\sum_{k_i} \chi_{k_i} c^\dag c \ket{k_i}\bra{k_i},
\end{equation}
where we make use of the following definitions, valid for a generic multi-qubit setup,
\begin{eqnarray}
J_{ij}^{(lm)}=\frac{g_ig_j(\omega_i+\omega_j+l\delta_i+m\delta_j-2\omega_r)}{2(\omega_i+l\omega_i-\omega_r)(\omega_j+m\delta_j-\omega_r)}\\
\chi_{k_i}=\frac{g_i^2(\delta_i-\omega_i+\omega_r)}{(\omega_i+k\delta_i-\omega_r)(\omega_i+(k-1)\delta_i-\omega_r)}\\
\tilde{\omega}_{k_i}=k\omega_i+\frac{\delta_i}{2}k(k-1)+\frac{kg_i^2}{\omega_i+(k-1)\delta_i-\omega_r}.
\end{eqnarray}
In the limit $|\tilde{\omega}_i-\tilde{\omega_j}|\gg J_{ij}^{(lm)}$, the qubit-qubit exchange interaction can be treated perturbatively with respect to the free energy term in Eq.~(\ref{DressedHam}), and results in a further dressing of the energy levels. Considering a four-qubit system, the resulting Hamiltonian on the qubit subspace can be modeled with a diagonal operator acting on four qubits
\begin{equation}
\label{Heff}
H_{\textrm{eff}}=\sum_{i=1}^4\Xi_p^{(i)}\sigma^z_i+\sum_{i\neq j} \zeta_p^{(ij)}\sigma^z_i \sigma^z_j +\sum_{a\neq b \neq c}\zeta_p^{(abc)}\sigma^z_a\sigma^z_b\sigma^z_c+\zeta_p^{4}\sigma^z_1\sigma^z_2\sigma^z_3\sigma^z_4,
\end{equation}
where the coefficients $\Xi_p,\zeta_p^{(ij)},\zeta_p^{(abc)},\zeta_p^{4}$ depend on the number of photons in the cavity $p$. To obtain them, we compute the corrections from the exchange process $J_{ij}^{(lm)}$ to the uncoupled level energies, $E_{1_jp}$, $E_{1_i1_jp}$, and $E_{1_i1_j1_lp}$ and $E_{1111p}$, representing one excitation on qubit $j$, two excitations distributed on qubits $i,j$, three excitations on qubits $a,b,c$ and the all-excited state, respectively. The first-order corrections are zero, while the second-order corrections can be written as
\begin{eqnarray}
E_{1_i1_jp}^{(2)}&=&E_{1_i1_jp}
+\sum_{k\neq i,j} \left(\frac{ J_{jk}^{(10)}}{E_{1_i1_jp}-E_{1_i1_kp}}
+ \frac{ J_{ik}^{(10)}}{E_{1_i1_jp}-E_{1_j1_kp}}\right)
+\left(\frac{ J_{ij}^{(21)}}{E_{1_i1_j p}-E_{2_jp}}+\frac{ J_{ji}^{(21)}}{E_{1_i1_jp}-E_{2_ip}}\right),\\
E_{1_i1_j1_lp}^{(2)}&=&E_{1_i1_j1_lp}
+\left(\frac{ J_{ik}^{(10)}}{E_{1_i1_j1_lp}-E_{1_{k}1_j1_lp}}+\frac{ J_{jk}^{(10)}}{E_{1_i1_j1_lp}-E_{1_i1_k1_lp}}+\frac{ J_{lk}^{(10)}}{E_{1_i1_j1_lp}-E_{1_i1_j1_kp}}\right)_{k\neq i,j}\\
&+&\Big(\frac{ J_{ij}^{(21)}}{E_{1_i1_j1_lp }-E_{2_j1_lp}}+\frac{ J_{il}^{(21)}}{E_{1_i1_j1_lp}-E_{1_j2_lp}}+\frac{ J_{ji}^{(21)}}{E_{1_i1_j1_lp}-E_{2_i2_lp}}+\frac{ J_{jl}^{(21)}}{E_{1_i1_j1_lp}-E_{1_i2_lp}}
+\frac{J_{li}^{(21)}}{E_{1_i1_j1_lp}-E_{2_i2_lp}}+\frac{J_{lj}^{(21)}}{E_{1_i1_j1_lp}-E_{1_i2_jp}}\Bigg),\nonumber\\
E_{1111p}^{(2)}&=&E_{1111p}+\sum_{a\neq b\neq c}\frac{J^{(21)}_{jk}}{E_{1111p}-E_{2_a1_b1_cp}}.
\end{eqnarray}
Therefore the effective Hamiltonian at second order in the perturbation can be written as
\begin{align}
H_{eff}^p=&\textrm{diag}\Big(E_{0000p}, E^{(2)}_{1000p},E^{(2)}_{0100p}, E^{(2)}_{0010p}, E^{(2)}_{0001p},E^{(2)}_{1100p},\\ &E^{(2)}_{0110p},E^{(2)}_{0011p},E^{(2)}_{1010p}, E^{(2)}_{0101p}, E^{(2)}_{1001p}, E^{(2)}_{1110p}, E^{(2)}_{0111p}, E^{(2)}_{1011p}, E^{(2)}_{1101p}, E^{(2)}_{1111p}\Big).  \nonumber
\end{align}
The coefficients in Eq.~(\ref{Heff}), can be obtained tracing  $\Xi_p^{(i)}=\textrm{Tr}[H_{eff}^p\sigma_i^z]/16$, $\zeta_p^{(ij)}=\textrm{Tr}[H_{eff}^p\sigma_i^z\sigma_{j}^z]/16$, $\zeta_p^{(abc)}=\textrm{Tr}[H_{eff}^p\sigma^z_a\sigma^z_b\sigma^z_c]/16$, and $\zeta^4_p=\textrm{Tr}[H_{eff}^p\sigma^z_1\sigma^z_2\sigma^z_3\sigma^z_4]/16$.
In the subspace with no photons in the resonator ($p=0$), we obtain
\begin{eqnarray}
\Xi_0^{(i)}&=&-\frac{1}{2}{\omega}_i+\frac{1}{4}\sum_{j\neq i}\left[\frac{\left[J_{ij}^{(01)}\right]^2}{\omega_i-\omega_j}+\frac{\left[J_{ij}^{(12)}\right]^2(\delta_i+\delta_j)}{4(\omega_i-\omega_j+\delta_i)(\omega_i-\omega_j-\delta_j)}\right],\\
\zeta_0^{(ij)}&=&\frac{\left[J_{ij}^{(12)}\right]^2(\delta_i+\delta_j)}{4(\omega_i-\omega_j+\delta_i)(\omega_i-\omega_j-\delta_j)},
\end{eqnarray}
while at second order in the energetic corrections there is no contribution from weight three and four operators to the dynamics, $\zeta_p^{(abc)}=0$, $\zeta^4_p=0$, for all cavity sectors $p$.
We now add the cavity dispersive interactions and a drive to the bus cavity $\epsilon(t)(c+c^\dag)$ to the Hamiltonian in Eq.~(\ref{Heff}), with $\epsilon(t)=\epsilon_I(t)\cos(\omega_d t)+\epsilon_Q(t)\sin(\omega_d t)$ being a drive at frequency $\omega_d$ and quadratures $\epsilon_I(t),\epsilon_Q(t)$. We obtain
\begin{eqnarray}
\label{Heffp}
H'_{\textrm{eff}}&=&\omega_r c^\dag c +\sum_{j,k,l,m} \chi_{jklm} c^\dag c \ket{jklm}\bra{jklm} +\sum_{i=1}\Xi_p^{(i)}\sigma^z_i+\sum_{i\neq j} \zeta_p^{(ij)}\sigma^z_i \sigma^z_j +\epsilon(t)(c+c^\dag),
\end{eqnarray}
where we have introduced a compact notation for the Stark shifts of all the qubits, $\chi_{jklm}=\chi_{j_1}+\chi_{k_2}+\chi_{l_3}+\chi_{m_4}$. We then apply the frame transformation
\begin{equation}
\label{frame}
R(t)=e^{-it\left(\sum_{i=1}^4\Xi_p^{(i)}\sigma^z_i+\omega_d c^\dag c\right)}
\end{equation}
to $H'_\textrm{eff}$, and make use of a generalized P-representation to encode the state of the cavity, writing the total density matrix for the qubit-cavity system as
\begin{align}
\label{rho}
\rho(t)&=\sum_{jklm} p_{jklm}\ket{\alpha_{jklm}(t),jklm}\bra{\alpha_{jklm}(t),jklm}\nonumber\\
&+ \sum_{jklm\neq abcd} \frac{e^{i\mu_{jklm,abcd}(t)}}{\langle\alpha_{abcd}(t)|\alpha_{jklm}(t)\rangle}\ket{\alpha_{abcd}(t),abcd}\bra{\alpha_{jklm}(t),jklm}.
\end{align}
We have defined $\alpha_{jklm}$ as a coherent state on the $jklm$ sector of the four qubits and $\tilde\Delta_{jklm}=i(\omega_r-\omega_d+\chi_{jklm})+\kappa/2$, $\kappa$ being the leakage rate of the cavity. The solution to a master equation
\begin{equation}
\dot\rho(t)=-i[R(t)H'_{\textrm{eff}}R^\dag (t),\rho(t)]+\kappa \mathcal{D}[c]\rho(t),
 \end{equation}
involving Hamiltonian in Eq.~(\ref{Heffp}) in the frame Eq.~(\ref{frame}) and a dissipative term for the cavity $\mathcal{D}[c]\rho(t)=(2c\rho c^\dag-c^\dag c \rho(t)-\rho (t) c^\dag c)/2$, can be expressed via the closed form
\begin{align}
\label{ClosedFormSolution1} \alpha_{jklm}(t)&=\alpha_{jklm}(0)e^{-\tilde{\Delta}_{jklm}t}-\frac{i}{2}\int_0^t e^{-\tilde{\Delta}_{jklm}(t-t')} \tilde{\epsilon}(t') dt',\\
\label{ClosedFormSolution2} \mu_{jklm,abcd}(t)&=\mu_{jklm,abcd}(0)+(\chi_{abcd}-\chi_{jklm})\int_0^t \alpha_{abcd}^*(t')\alpha_{jklm}(t') dt' + \zeta_{jklm,abcd}.
\end{align}
In the above expression $\tilde\epsilon (t)=\epsilon_I(t)+i\epsilon_Q(t)$ and we have defined the contribution from the static interactions to the dynamics
\begin{align}
\zeta_{jklm,abcd}=&\big[-i\zeta^{(12)}_0((-1)^{j+k}-(-1)^{a+b})-i\zeta^{(13)}_0((-1)^{j+l}-(-1)^{a+c})\nonumber\\
&-i\zeta^{(14)}_0((-1)^{j+m}-(-1)^{a+d})-i\zeta^{(23)}_0((-1)^{k+l}-(-1)^{b+c})\nonumber\\
&-i\zeta^{(24)}_0((-1)^{k+m}-(-1)^{b+d})-i\zeta^{(34)}_0((-1)^{l+m}-(-1)^{c+d})\big],
\end{align}
assuming that the dynamics does not involve population of higher levels of the cavity, so that $\zeta_p^{(ij)}\approx\zeta_0^{(ij)}$.
A diagonal unitary operator $U(t)=\textrm{diag}(e^{i \theta_{0000}(t)},e^{i \theta_{1000}(t)},...e^{i \theta_{1111}(t)})$ induces the transformation on the density matrix of the system $e^{i(\theta_{jklm}(t)-\theta_{abcd}(t))}\rho_{jklm,abcd}$. From there one can find $\mu_{jklm,abcd}(t)=\theta_{jklm}(t)-\theta_{abcd}(t)$.

Therefore, from the solution in Eqs.~(\ref{ClosedFormSolution1}-\ref{ClosedFormSolution2}) one can retrieve the dynamical rates of the $\sigma^z_i\sigma^z_j$, $\sigma^z_a\sigma^z_b\sigma^z_c$, $\sigma^z_1\sigma^z_2\sigma^z_3\sigma^z_4$ contributions to the dynamics of the system. This can be obtained by taking the traces $\theta_{ij}(t)=\textrm{Tr}[-i\log(U(t))\sigma^z_i\sigma^z_j]/16$, $\theta_{abc}(t)=\textrm{Tr}[-i\log(U(t))\sigma^z_a\sigma^z_b\sigma^z_c]/16$, $\theta_4(t)=\textrm{Tr}[-i\log(U(t))\sigma^z_1\sigma^z_2\sigma^z_3\sigma^z_4]/16$, and rewriting them in terms of the $\mu_{jklm,abcd}$ phases, obtained from Eq.~(\ref{ClosedFormSolution2}). For example, for the interactions between the first and second qubit, one has
\begin{align}
\label{Z1Z2Expr}
\theta_{Z_1Z_2}(t)&=\frac{1}{16}\textrm{Tr}[-i\log(U(t))Z_1Z_2]=\frac{1}{16} \Big(\theta_{0000}(t)-\theta_{0001}(t)-\theta_{0010}(t)+\theta_{0011}(t)+\theta_{0100}(t)
-\theta_{0101}(t)-\theta_{0110}(t)\nonumber\\
&+\theta_{0111}(t)+\theta_{1000}(t)-\theta_{1001}(t)-\theta_{1010}(t)+\theta_{1011}(t)
   +\theta_{1100}(t)+\theta_{1101}(t)-\theta_{1110}(t)+\theta_{1111}(t)\Big)\nonumber\\
   &=\frac{1}{16}\Big(\mu_{0000,0001}(t)+\mu_{0011,0010}(t)+\mu_{0100,0101}(t)+\mu_{0111,0110}(t)
   +\mu_{1000,1001}(t)+\mu_{1011,1010}(t)\nonumber\\
   &+\mu_{1100,1101}(t)+\mu_{1111,1110}(t)\Big).
\end{align}
To understand the magnitudes of the various interactions that appear, we can compute their value when the bus cavity is driven with a constant unmodulated tone $\tilde{\epsilon}(t)=\tilde{\epsilon}_0$. In this case, from Eq.~(\ref{ClosedFormSolution1}), one has a steady state bus response of $\alpha_{jklm}=-i\tilde{\epsilon}_0/2\tilde{\Delta}_{jklm}$. Assuming identical qubits (i.e. same frequencies, anharmonicities and Stark shifts $\chi$), the dynamical rates for two, three and four body terms are found to be
\begin{eqnarray}
\dot{\theta}^{s.s.}_{Z_iZ_j}&=&-\frac{|\tilde{\epsilon}_0|^2\chi^2}{8 \Delta(\Delta+2\chi)(\Delta+4\chi)},\label{theta2}\\
\dot{\theta}^{s.s.}_{Z_aZ_bZ_c}&=&-\frac{3|\tilde{\epsilon}_0|^2\chi^3}{16\Delta(\Delta+\chi)(\Delta+3\chi)(\Delta+4\chi)},\label{theta3}\\
\dot{\theta}^{s.s.}_{Z_1Z_2Z_3Z_4}&=&-\frac{3|\tilde{\epsilon}_0|^2\chi^4}{8\Delta(\Delta+\chi)(\Delta+2\chi)(\Delta+3\chi)(\Delta+4\chi)}.\label{theta4}
\end{eqnarray}
In the limit in which the Stark shifts are much smaller that the drive-cavity detuning $\Delta=\omega_d-\omega_r$, $\chi\ll\Delta$, there is a clear scaling of these rates with powers of $\chi/\Delta$. In fact, by inspection $\dot\theta_{Z_aZ_bZ_c}^{s.s.}\propto(\chi/\Delta)\dot\theta_{Z_iZ_j}^{s.s.}$, $\dot\theta_{Z_1Z_2Z_3Z_4}^{s.s.}\propto(\chi/\Delta)^2\dot\theta_{Z_iZ_j}^{s.s.}$. Therefore, contributions to the driven dynamics will be given mainly by weight-two Pauli terms, and higher order interactions will be slow and increasingly hard to measure. The off-diagonal matrix elements of $\rho(t)$ are affected by an induced dephasing from driving of the bus cavity. To estimate the rate of this coherence loss, we can consider the fastest coherence decay between the states $\ket{0000}$ and $\ket{1111}$, happening at a rate given by
\begin{equation}
\textrm{Im}[\dot{\mu}_{0000,1111}]=\frac{2\kappa|\tilde{\epsilon}_0|^2\chi^2}{\Delta^2(\Delta+4\chi)^2}.
\end{equation}
In the limit $\chi\ll\Delta$, the dephasing rate is proportional to the steady state rates
\begin{eqnarray}
\textrm{Im}[\dot{\mu}_{0000,1111}]&\propto&\dot{\theta}^{s.s.}_{Z_iZ_j}\frac{\kappa}{\Delta}, \\
\textrm{Im}[\dot{\mu}_{0000,1111}]&\propto&\dot{\theta}^{s.s.}_{Z_aZ_bZ_c}\frac{\Delta}{\chi}\frac{\kappa}{\Delta}=\dot{\theta}^{s.s.}_{Z_aZ_bZ_c}\frac{\kappa}{\chi},\\
\textrm{Im}[\dot{\mu}_{0000,1111}]&\propto&\dot{\theta}^{s.s.}_{Z_1Z_2Z_3Z_4}\left(\frac{\Delta}{\chi}\right)^2\frac{\kappa}{\Delta}=\dot{\theta}^{s.s.}_{Z_1Z_2Z_3Z_4}\frac{\Delta\kappa}{\chi^2}.
\end{eqnarray}
It is therefore possible to find a parameter regime in which the dephasing is negligible with respect to the entangling rate $\dot{\theta}^{s.s.}_{Z_iZ_j}$. For example, by choosing large detunings $\Delta$ and compensating for this choice with high driving power, one can effectively suppress the dephasing relative to the gate rate. However, finding such a parameter range for $\dot\theta^{s.s.}_{Z_aZ_bZ_c}$ and $\dot\theta^{s.s.}_{Z_1Z_2Z_3Z_4}$ is much harder because typically $\chi\ll\Delta$.

\subsection{Measurements of phase rates\label{PhaseRates}}

In this section we give additional details about the measurement of the phase interactions that we perform. To verify the behavior of the multiqubit experiment against the solution of Eqs.~(\ref{ClosedFormSolution1},\ref{ClosedFormSolution2}), we first perform a set of Ramsey experiments on the four different qubits. The $T_2^*$ decoherence time of the qubits is taken into account in our model by adding pure imaginary terms to the off-diagonal density matrix elements in Eq.~(\ref{ClosedFormSolution2}),
\begin{eqnarray}
\label{mudecoh}
\mu^{(d)}_{jklm,abcd}(t)&=&\mu_{jklm,abcd}(t)+\mu^{T_2^*}_{jklm,abcd}(t),\\
\mu^{T_2^*}_{jklm,abcd}(t)&=&i(|j-a|t/T_{2}^{*(1)}+|k-b|t/T_{2}^{*(2)}+|l-c|t/T_{2}^{*(3)}+|m-d|t/T_{2}^{*(4)})
\end{eqnarray}
To measure a single $\mu(t)$ phase, we perform a Ramsey experiment on each one of the qubits in the system. Focusing on qubit 1, we first prepare the system in the superposition $(\ket{0000}+i\ket{0001})/\sqrt{2}$, starting from the $\ket{0000}$ state and applying a $\pi/2$ pulse around the $X$ axis on the first qubit. Then, a RIP tone is applied to the cavity, inducing a relative phase on the state $(e^{i\theta_{0000}(t)}\ket{0000}+ie^{i\theta_{0001}(t)}\ket{0001})/\sqrt{2}\equiv\ket{0000}+ie^{i\mu_{0001,0000}(t)}\ket{0001})/\sqrt{2}$. Under the action of the RIP tone, and taking into account decay and dephasing rates, the density matrix on the reduced subspace of the first qubit can be written as
\begin{equation}
\label{rhoRamsey}
\rho_1(t)=\left(
\begin{array}{cc}
\frac{1}{2}e^{-t/T_{1}^{(1)}} & ie^{i\mu^{(d)}_{0001,0000}(t)}\\
 -ie^{-i\mu^{*(d)}_{0001,0000}(t)} & 1-\frac{1}{2}e^{-t/T_{1}^{(1)}} \\
\end{array}\right),
\end{equation}
where we have used the definition for the off-diagonal matrix element, that includes $T_{2}^*$ coherence times as in Eq.~(\ref{mudecoh}). One can then apply a $\pi/2$ rotation about the $Y$ axis, and measure the probability of finding the qubit in the excited and ground state, given by
\begin{eqnarray}
\label{RamseyProb1}
P_\uparrow(t)&=&\frac{1}{4}\left[2+i\left(e^{i\mu^{(d)}_{0001,0000}(t)}-e^{-i\mu^{*(d)}_{0001,0000}(t)}\right)\right]=\frac{1}{2}\left[1-\sin(\mu_{0001,0000}(t))e^{-t/T_2^{(1)}}\right],\\
\label{RamseyProb2}
P_\downarrow(t)&=&1-P_\uparrow (t).
\end{eqnarray}
The photonic population of the cavity can be obtained by taking into account that the RIP interactions acts while the qubit state is in the state described by Eq.~(\ref{rhoRamsey}). The weighted superposition of states $\ket{0000}$ and $\ket{0001}$ generates the corresponding superposition of cavity states, according to Eq.~(\ref{ClosedFormSolution1}),
\begin{equation}
\langle n(t) \rangle=\frac{1}{2}\left[1-\sin(\mu_{0001,0000}(t))\right]|\alpha_{0000}(t)|^2+\frac{1}{2}\left[1+\sin(\mu_{0001,0000}(t))\right]|\alpha_{0001}(t)|^2
\end{equation}
The presence of photons at the end of the Ramsey protocol signals a non-adiabatic behavior of the RIP interactions. Measured Ramsey fringes for the four qubits in device A are shown in Fig.~\ref{SupplRamsey} and compared to the corresponding theoretical prediction obtained by using the expressions in Eq.~(\ref{RamseyProb1}) and the solution of Eqs.~(\ref{ClosedFormSolution1}-\ref{ClosedFormSolution2}).
\begin{figure}
\includegraphics[width=7in]{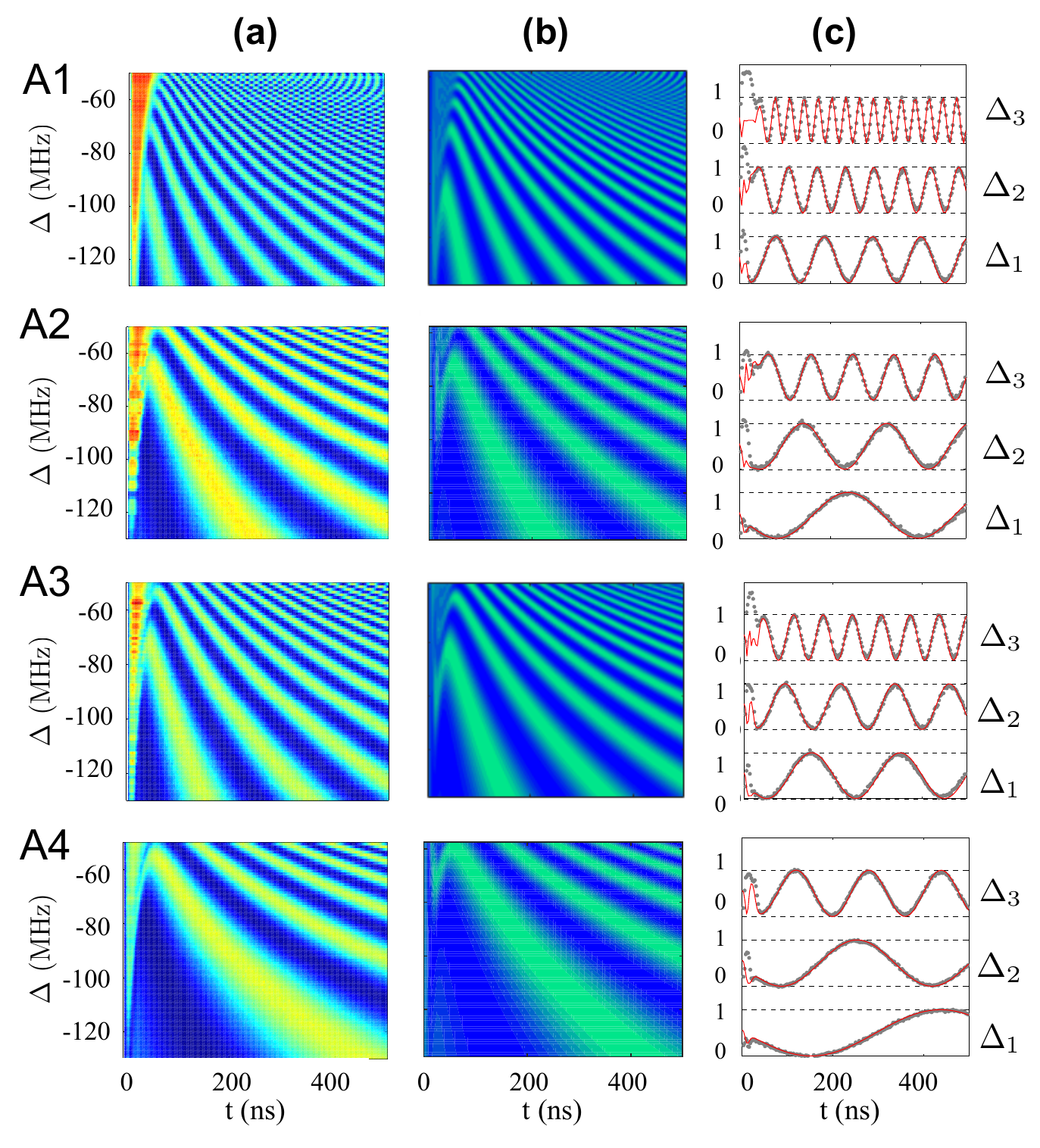}
\caption{\label{SupplRamsey} Ramsey fringe measurements on the four qubits A1-A4 for the full RIP interaction. (a) Measured experimental probability $P_\uparrow$ of finding the specified qubit in the $A$ setup in the excited state after the Ramsey protocol, as a function of the drive frequency $\omega_d$. (b) Corresponding theory prediction, obtained using expression in Eq.~(\ref{RamseyProb1}), using pulse amplitudes $\tilde{\epsilon}_R/2\pi=262,258,283,252$~MHz, for qubit $A1$, $A2$, $A3$, and $A4$ respectively. (c) Ramsey fringes at specified drive frequencies. The theory prediction is obtained by fitting with the drive amplitudes used for (b), adding relative power corrections of $\sim\pm0.05\tilde{\epsilon}_R$, for detunings of $\Delta_1/2\pi=-137$~MHz, $\Delta_2/2\pi=-119$~MHz, $\Delta_3/2\pi=-97$~MHz. }
\end{figure}

\subsection{Echoed Phase Interactions and GHZ sequence}
\begin{figure*}
\includegraphics[width=7in]{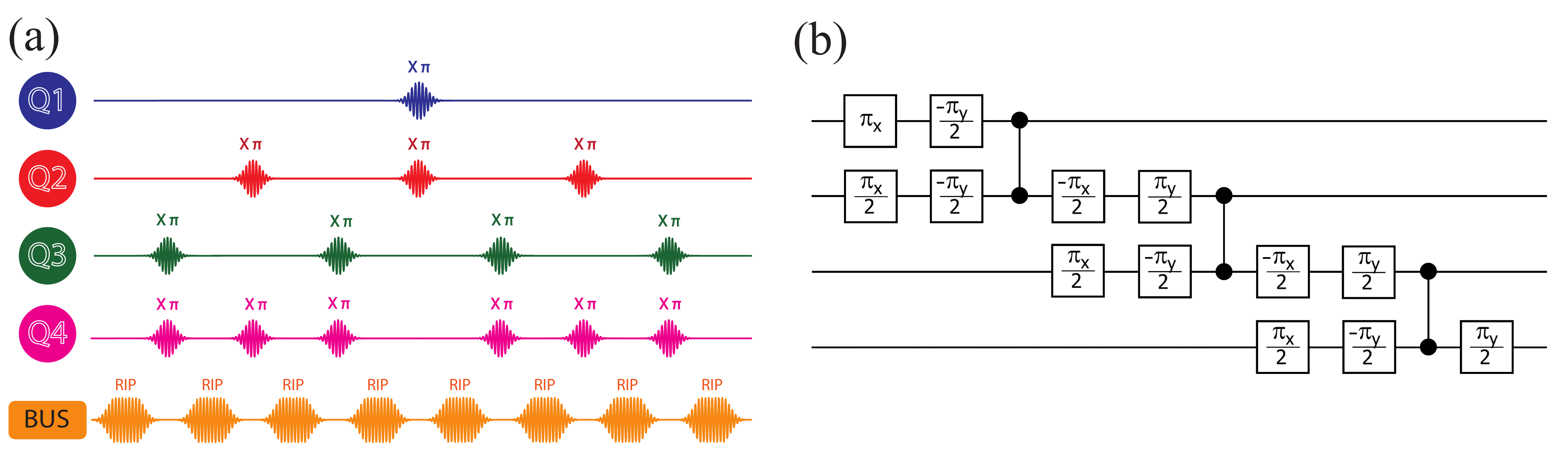}
\caption{\label{EchoSchemes} (a) The 8-pulse refocused RIP gate scheme for ZZZZ. (b) Pulse sequences to create a 4-qubit GHZ state. }
\end{figure*}

To isolate Pauli operations of weight two, three and four out of the whole set of phase interactions that take place as in Eq.~(\ref{Heffp}), we perform echo sequences, reported in Fig.~2 in the main text for $\sigma_2^z\sigma_3^z$ and $\sigma_1^z\sigma_2^z\sigma_3^z$  and in Fig.~\ref{EchoSchemes} here for the $\sigma^z_1\sigma^z_2\sigma^z_3\sigma^z_4$ interaction.
Ramsey experiments are then performed in order to measure the rate of each process, following the discussion in section~\ref{PhaseRates}.
For example, to measure the $\sigma_1^z\sigma_2^z$ term, we prepare the state in the $(\ket{0000}+i\ket{0001})/\sqrt{2}$ superposition with a $\pi/2$ rotation on qubit 1 about the $X$ axis, and interleave RIP gate tones with single qubit pulses. For the sequence that echoes the $\theta_{Z_{1}Z_{2}}$ term, performing the echo steps leads to the phase accumulation
\begin{align}
&\frac{1}{\sqrt{2}}\Big(e^{i\left(\theta_{0000}(t)+\theta_{0011}(t)+\theta_{0111}(t)+\theta_{0100}(t)+\theta_{1100}(t)+\theta_{1111}(t)+\theta_{1011}(t)+\theta_{1000}(t)\right)}\ket{0000}\\
&+ie^{i\left(\theta_{0001}(t)+\theta_{0010}(t)+\theta_{0110}(t)+\theta_{0101}(t)+\theta_{1101}(t)+\theta_{1110}(t)+\theta_{1010}(t)+\theta_{1001}(t)\right)}\ket{0001}\Big)\\
&=\frac{1}{\sqrt{2}}\Big(\ket{0000}+ie^{-i\left(\mu_{0000,0001}(t)+\mu_{0011,0010}(t)+\mu_{0100,0101}(t)+\mu_{0111,0110}(t)
   +\mu_{0000,0001}(t)+\mu_{1011,1010}(t)+\mu_{1100,1101}(t)+\mu_{1111,1110}(t)\right)}\ket{0001}\Big)\\
   &\equiv\frac{1}{\sqrt{2}}\Big(\ket{0000}+ie^{-i\mu_{Z_1Z_2}(t)}\ket{0001}\Big).
\end{align}
where $t$ is the interaction time of each RIP gate.
Notice that the relative phase coincides with the expression in Eq.~(\ref{Z1Z2Expr}). An additional $\pi/2$ rotation on qubit 1 about the $Y$ axis can map the $\theta_{Z_1Z_2}$ phase onto $P_\uparrow, P_\downarrow$, following Eqs.~(\ref{RamseyProb1},\ref{RamseyProb2}), taking into account this time that the decoherence is acting for a time $8t+7 \tau$, where $\tau\approx 37$~ns is the time for a single qubit $\pi$ pulse.
To compute the residual photon population of the cavity at time $t$, used to produce the inset in the main text in Fig.~2(a), one can use Eq.~(\ref{ClosedFormSolution1}) for each echo step, using as initial condition the final photon population of the previous step,
\begin{equation}
\langle n(t) \rangle=\frac{1}{2}\left[1+\sin(\mu_{Z_1Z_2}(t))\right]|\alpha^{(0)}_{Z_1Z_2}(t)|^2+\frac{1}{2}\left[1-\sin(\mu_{Z_1Z_2}(t))\right]|\alpha^{(1)}_{Z_1Z_2}(t)|^2,
\end{equation}
where $\alpha^{(0)}_{Z_1Z_2}(t)$ and $\alpha^{(1)}_{Z_1Z_2}(t)$ are the photon population at the end of the RIP echo sequence of Fig.~\ref{EchoSchemes}, starting from the state $\ket{0000}$ or $\ket{0001}$, respectively. To calculate them we make use of the closed form in Eq.~(\ref{ClosedFormSolution1}) for each time interval where the RIP tone is applied to the cavity, taking into account the corresponding $jklm$ qubit state, that depends on the specific $\pi$-pulse sequence used. Therefore, a system of nested equations is defined, connecting the time-dependent cavity state at each RIP step. For example, $\alpha^{(0)}_{Z_1Z_2}(t)$ can be calculated as
\begin{equation}
\begin{cases}
\alpha^{(8)}_{1101}(t)&=\alpha^{(7)}_{1111}(t)e^{-\tilde{\Delta}_{1101}t}-\frac{i}{2}\int_0^t e^{-\tilde{\Delta}_{1101}(t-t')} \tilde{\epsilon}(t') dt'\equiv\alpha^{(0)}_{Z_1Z_2}(t)\\
\alpha^{(7)}_{1111}(t)&=\alpha^{(6)}_{0011}(t)e^{-\tilde{\Delta}_{1111}t}-\frac{i}{2}\int_0^t e^{-\tilde{\Delta}_{1111}(t-t')} \tilde{\epsilon}(t') dt'\\
&\vdots\\
\alpha^{(1)}_{0000}(t)&=-\frac{i}{2}\int_0^t e^{-\tilde{\Delta}_{0000}(t-t')} \tilde{\epsilon}(t') dt',
\end{cases}
\end{equation}
where $t$ is the duration time of the single RIP tones in every echo step. A similar system of equations can be used to obtain $\alpha^{(1)}_{Z_1Z_2}(t)$.

With CZ gates constructed from the 8-pulse refocused RIP gate scheme in Fig.2 in the main text, we generate a 4-qubit GHZ state using the pulse sequence shown in Fig.~\ref{EchoSchemes}(b).

\subsection{Measurement-Induced Dephasing}

To compute the measurement-induced dephasing in Fig. 3(c) in the main text, we use a total pulse time of $2 \times \tilde{\tau}_g = 2 \times 266.7$~ns$ = 533$~ns, and calibrate numerically the drive amplitude $\tilde{\epsilon}_A$ needed to perform a $\pi/2$ ZZ gate, using parameters for qubits A2 and A3, as a function of drive detuning $\Delta$. Once the drive amplitude $\tilde{\epsilon}_A$ is calibrated for each detuning of Fig.~3(c), we numerically compute the decay of the matrix elements $\rho_{0000,0001},\rho_{0000,0010},\rho_{0000,0100},\rho_{0000,1000}$ under the action of the driving pulse $\tilde{\epsilon}_R(t)=\tilde{\epsilon}_A (1+\cos(\pi\cos(\pi t /\tilde{\tau})))$, using Eq.~(\ref{ClosedFormSolution2}). We fit the decay of each off-diagonal matrix element to the model $\rho(t)=\exp(-t/T_2)$, obtaining effective $T_2$ decoherence times for each qubit, which we use to compute the contribution from measurement-induced dephasing to the gate fidelities in Fig.~3(c).
\end{widetext}

\end{document}